\documentclass[prl,twocolumn,showpacs, amssymb, amsmath, aps, superscriptaddress] {revtex4}

\usepackage{graphicx}
\usepackage{dcolumn}
\usepackage{bm}
\usepackage{epstopdf}
\usepackage{latexsym}
\usepackage{mathrsfs}
\begin{document}


\title{Accurate Microwave Control and Real-Time Diagnostics of Neutral Atom Qubits}

\author{Worawarong Rakreungdet}
\affiliation{College of Optical Sciences, University of Arizona, Tucson, AZ 85721}
\author{Jae Hoon Lee}
\affiliation{College of Optical Sciences, University of Arizona, Tucson, AZ 85721}
\author{Kim Fook Lee}
\affiliation{Department of Physics, Michigan Technological University, Houghton, MI 49931}
\author{Brian E. Mischuck}
\affiliation{Department of Physics and Astronomy, University of New Mexico,Albuquerque, NM 87131}
\author{Enrique Montano}
\affiliation{College of Optical Sciences, University of Arizona, Tucson, AZ 85721}
\author{Poul S. Jessen}
\affiliation{College of Optical Sciences, University of Arizona, Tucson, AZ 85721}

\date{\today}

\begin{abstract}
We demonstrate accurate single-qubit control in an ensemble of atomic qubits trapped in an optical lattice.  The qubits are driven with microwave radiation, and their dynamics tracked by optical probe polarimetry.  Real-time diagnostics is crucial to minimize systematic errors and optimize the performance of single-qubit gates, leading to fidelities of $0.99$ for single-qubit $\pi$ rotations.  We show that increased robustness to large, deliberately introduced errors can be achieved through the use of composite rotations.  However, during normal operation the combination of very small intrinsic errors and additional decoherence during the longer pulse sequences precludes any significant performance gain in our current experiment.  \end{abstract}


\maketitle

\section{I. Introduction}
Optically trapped neutral atoms have been widely considered as a physical platform for the implementation of quantum information processing.  Qubits encoded in the hyperfine ground manifold offer long coherence times and are readily manipulated using the tools of laser spectroscopy, whereas the two-qubit interactions required for entangling quantum gates are more difficult to implement due to the weak coupling between atoms in the electronic ground state. For this reason, two-qubit quantum logic is generally regarded as the core challenge for the neutral atom platform, and most theoretical and experimental efforts have focused on developing viable gate schemes based on e. g. cold collisions \protect\cite{Brennen1999}\protect\cite{Jaksch1999}\protect\cite{Mandel2003}\protect\cite{Anderlini2007} or dipole-dipole interactions between Rydberg states \protect\cite{Jaksch2000}\protect\cite{Lukin2001}\protect\cite{Urban2008}\protect\cite{Gaetan2008}.  From these first-generation experiments it is clear, however, that accurate single-qubit manipulation is already critical and not always trivial to achieve in a laboratory setting. In practice, noisy or inaccurate driving fields and trap potentials, along with other perturbations, cause unpredictable errors in the effective Rabi frequency and detuning for the qubit two-level system, and reduce the fidelity of rotations on the Bloch sphere.  In this situation it is useful to explore both how far one can reasonably go through a direct effort to eliminate these sources of error, and how one can use composite pulse techniques that are robust against errors as is routinely done by the NMR community \protect\cite{Cummins2003}\protect\cite{chuangrevmod}, and more recently for qubits encoded in trapped ions \protect\cite{Gulde2003}\protect\cite{Timoney2008}, electron spins \protect\cite{Morton2005} and superconducting circuits \protect\cite{Collin2004}.  Ultimately, very high fidelity single-qubit operations can become an important resource to combat errors and dephasing across the qubit ensemble \protect\cite{Viola1999}, through spin echoes and refocusing pulses as used in NMR quantum computation \protect\cite{chuangrevmod}\protect\cite{Vandersypen2001}.  It is likely that trapped atom implementations will need to adopt a similar strategy.

For ensembles of atomic qubits trapped in optical lattices, precise single-qubit control is compromised by spatial variations in the lattice potential and the associated inhomogeneous broadening of the AC Stark shifted qubit transition frequency.  Furthermore, some collisional gate schemes require that a qubit be encoded in magnetic field-sensitive states, such as $ \vert 0 \rangle = \vert F=3,m_f=3 \rangle$ and $ \vert 1 \rangle = \vert F=4,m_f=4 \rangle$ in the case of the Cs atom qubits used in our work here, in order to permit state-dependent motion in the optical potential and/or to suppress spin changing collisions \protect\cite{Jaksch1999}.  In that case imperfect control over the magnetic field can easily become an important source of error.  In this letter we demonstrate that microwave-driven single-qubit gates with fidelities as high as $0.99$ can be achieved in parallel on $\sim10^7$  atomic qubits, by carefully optimizing the spatial and temporal uniformity of the optical lattice and of the applied and ambient magnetic fields. In doing so, a critical element has been the development of a non-perturbing method to observe qubit ensemble dynamics in real time and with good signal-to-noise ratio (SNR), based on optical probe polarimetry.  We further show that composite pulse techniques can significantly increase robustness against pulse imperfections and ensemble inhomogeneities.  One pulse sequence in particular, known as a rotary echo, greatly reduces sensitivity to these errors and allows easy observation of the coherence time in our system.  We expect that increased microwave irradiance, in combination with decreased scattering of lattice photons and the use of more advanced pulse techniques \protect\cite{Khaneja2004}, will allow us to improve the fidelity of single-qubit operations by at least another order of magnitude.

The remainder of this paper is organized as follows.  In Sec. II we describe our experimental apparatus, including our programmable microwave source and optical probe setup.  Sec. III discusses Rabi oscillations and the fidelity of simple rotations, Sec. IV describes the use of rotary echoes to extract coherence times, and Sec. V discusses the use of composite pulses to improve gate fidelities in the presence of errors. We summarize our findings and discuss the outlook for robust control of neutral atom qubits in Sec. VI.

\section{II. Experimental Setup}
\begin{figure}
[t]\resizebox{8.75cm}{!}
{\includegraphics{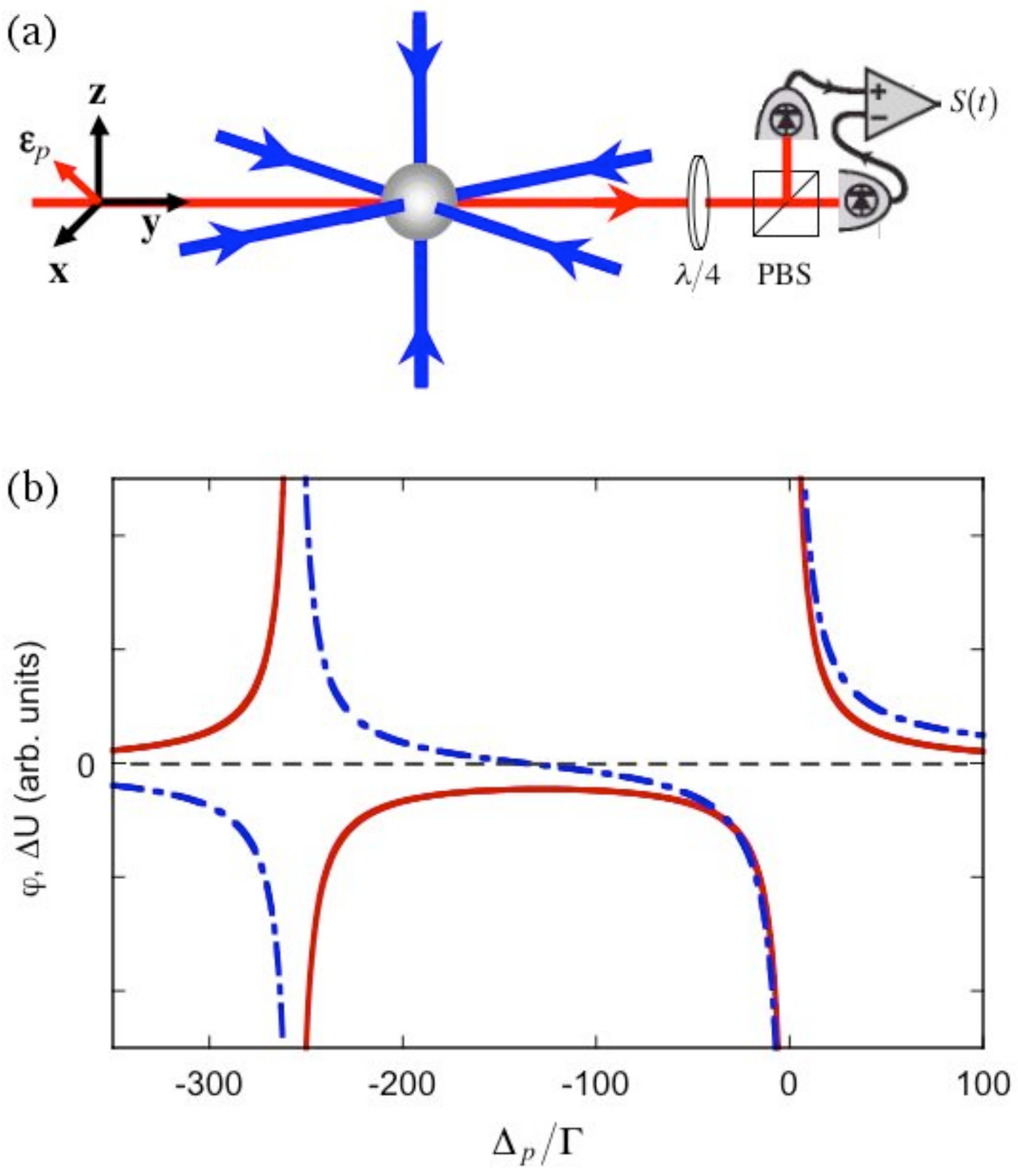}}
\caption{\label{fig:setup}(Color online) (a) Experimental setup, showing the optical lattice and the configuration used for optical probe polarimetry.   (b) Birefringent phase shift $\varphi$ seen by the probe (solid line), and the probe induced differential light shift $\Delta U$ of the qubit states (dashed line), as function of the probe detuning from the $6S_{1/2}(F=4)\rightarrow6P_{1/2}(F'=4)$ transition of the D1 line.}
\end{figure}

Our basic experimental setup is shown in Fig.~\ref{fig:setup}(a).  We begin by preparing a sample of $\sim10^7$  laser cooled Cs atoms in a cloud with a diameter of  $\sim0.8mm$.  The atoms are loaded into a three-dimensional optical lattice tuned $130GHz$  to the blue of the $6S_{1/2}(F=4)\rightarrow6P_{3/2}(F'=5)$  transition of the Cs D2 resonance line, by ramping up the lattice potential while the atoms are being simultaneously cooled by a near-resonant 3D optical molasses, continuing to cool the atoms for a few ms while in the lattice, and then extinguishing the optical molasses beams.  The optical lattice is formed by three independent, linearly polarized standing waves whose frequencies differ by at least $10MHz$ in order to eliminate interference between them.  Typical lattice depths are $U_L\sim100E_R$ , and typical vibrational frequencies in the lattice potential wells are $\omega_L\sim20E_R/\hbar\sim2\pi\times40kHz$, where $E_R=(\hbar k)^2/2M$  is the single photon recoil energy.   Once trapped in the lattice the atoms are optically pumped, producing a $~90\%$ population of the $ \vert 0 \rangle = \vert F=3,m_f=3 \rangle$ qubit state.  No additional cooling of the center-of-mass motion is performed, and measurements of the kinetic temperature indicate a mean vibrational excitation $\bar{n}\sim1$  for each degree of freedom.   At this point the qubits can either be kept in the lattice, or released into free fall to eliminate lattice-induced light shifts of the qubit transition frequency. 

Single-qubit rotations are implemented in parallel across the entire ensemble by illuminating the atoms with microwave radiation.  We use an inexpensive programmable microwave source consisting of an HP8672A synthesized signal generator running at $9.2GHz$, mixed with a $\sim30MHz$ signal from an SRS DS345 synthesized arbitrary waveform generator.  The DS345 provides both frequency fine-tuning and arbitrary phase modulation ($0.001^{\circ}$ resolution) under control of an uploaded computer generated waveform, and, if desired, can be amplitude modulated under control of an arbitrary analog waveform generated by, e. g., a second DS345.  By driving the qubits with a modulation sideband we thus achieve the precise, programmable phase control required to implement composite pulses. A digital switch inserted between the $9.2GHz$ local oscillator and the mixer allows the mixer output to be turned on/off with a transition time of $\sim3ns$.  The output from the mixer is amplified to $\sim2W$ by a solid state power amplifier, and radiated by a gain horn located approximately $20cm$ from the atomic sample. Due to reflections from various metallic surfaces in our apparatus we have only moderate control over the microwave polarization, but transitions to hyperfine states outside the qubit basis are Zeeman shifted out of resonance by a $\sim400mG$ bias magnetic field and effectively suppressed.  With this setup we can achieve resonant Rabi frequencies as high as $40kHz$ when driving the $\vert F=3,m_f=3 \rangle\rightarrow\vert F=4,m_f=4 \rangle$ transition, which corresponds to the largest magnetic dipole moment.

High fidelity rotations of the qubit pseudospin require precise knowledge of the resonant Rabi frequency $\chi$ and the effective detuning $\Delta$ of the microwave driving field relative to the qubit transition frequency.  In our setup $\chi$ varies slightly across the ensemble, due in part to the spatially inhomogeneous radiation pattern from the gain horn, and in part due to interference with microwave reflections.  Variations in $\Delta$ are caused by inhomogeneous broadening of the qubit transition frequency, which is light shifted by the spatially inhomogeneous lattice light field, and Zeeman shifted by the applied and ambient magnetic fields. For ensembles in far detuned lattices, the inhomogeneous spreads in $\chi$ and $\Delta$ easily become the most important factors limiting the fidelity of single-qubit rotations, and they must be carefully minimized to achieve good gate performance.  We accomplish this by driving Rabi oscillations between the qubit states, and by carefully aligning our lattice beams and aiming our microwave horn to maximize both the Rabi frequency and the decay time for the Rabi oscillation amplitude.  In doing so, an essential requirement is to have good diagnostic tools so that adjustments can be made and improvements detected as near as possible in real time.

We perform real-time, minimally perturbing measurements on our qubit ensemble by optical probe polarimetry, as described in detail in \protect\cite{Chaudhury2006}.  In this measurement scheme we choose the atomic quantization axis along the (vertical) $z$-axis, pass a weak, off-resonance probe beam through the ensemble along the $y$-axis, and choose the probe polarization to be linear and oriented at $45^{\circ}$ in the $x$-$z$ plane.  The probe frequency is tuned in-between the $6S_{1/2}(F=4)\rightarrow6P_{1/2}(F'=3,4)$ transitions of the D1 line, where the tensor polarizability is substantial for atoms in the $\vert1\rangle=\vert F=4,m_f=4\rangle$ qubit state.  In this situation the ensemble becomes birefringent by an amount proportional to the population of the logical-$\vert1\rangle$  state, and the probe polarization acquires a small degree of ellipticity, which is easily detected with a shot-noise limited polarimeter.  As demonstrated previously for the states involved in the atomic clock transition \protect\cite{Chaudhury2006}, any probe-induced, spatially inhomogeneous light shifts of the qubit transition frequency can be minimized by tuning the probe frequency to the appropriate ÒmagicÓ value, in this case $135$ natural linewidths ($633MHz$) to the red of the $F=4\rightarrow F'=4$ transition (Fig.~\ref{fig:setup}(b)).   In principle one can also detect the Faraday rotation caused by qubits in the logical-$\vert1\rangle$  state, a variation of probe polarimetry which is perhaps more familiar from its widespread use in magnetometry \protect\cite{Budker2002} and in measurements on hyperfine spin ensembles \protect\cite{Kuzmich2000}\protect\cite{Julsgaard2001}\protect\cite{Smith2004}.  However, for our qubit encoding the Faraday signal almost vanishes at the magic probe frequency and is therefore not as useful.  Overall, our measurement scheme provides high bandwidth access to the qubit dynamics, with little or no perturbation except for optical pumping caused by the scattering of probe photons.  Even for our relatively dilute atomic samples, the SNR's needed to align the lattice and optimize our apparatus for minimal qubit transition frequency broadening can be achieved in seconds.  Much higher SNR's are needed to measure the fidelity of highly accurate qubit rotations, but still require no more than a few minutes averaging.  As a consistency check or when more detailed information is needed, we complement the optical probe measurement with Stern Gerlach measurements \protect\cite{Klose2001} performed at discrete moments in time.  While time consuming, these provide access to the individual populations of all magnetic sublevels in the entire ground hyperfine manifold, and provide an absolute calibration for the initial preparation of the qubits by optical pumping.

\section{III. Rabi Oscillations and Simple Rotations}
An arbitrary qubit rotation $R_\mathbf{u}(\theta)$ can be specified by the axis $\mathbf{u}$ and angle $\theta$ of rotation.  The fidelity of a laboratory implementation of $R_\mathbf{u}(\theta)$ can be tested by initializing an ensemble of qubits in a state $\vert\psi_i\rangle$, applying the microwave control pulse, and measuring the absolute square of the overlap, $\mathscr{F}=\vert\langle\psi_{exp}\vert R_\mathbf{u}(\theta)\vert\psi_i\rangle\vert^2$, between the target state $R_\mathbf{u}(\theta)\vert\psi_i\rangle$ and the state $\vert\psi_{exp}\rangle$ produced in the experiment.  In general, one should sample the fidelity for a distribution of initial states, but for simple (as opposed to composite) rotations the fidelity depends only on the errors in $\mathbf{u}$ and $\theta$, and we can choose $\vert\psi_i\rangle=\vert0\rangle$ without loss of generality.  Specializing further to rotations by an angle $\theta=\pi$ around an axis in the equatorial plane of the Bloch sphere (a  $\pi$-gate), the fidelity can be determined simply by measuring the populations of the logical-$\vert0\rangle$ or -$\vert1\rangle$ states.  In practice, accurate determination of near-unit fidelities is best done by measuring the cumulative error in many successive applications of the rotation.  Experimentally, this is equivalent to driving Rabi oscillations and observing their decay due to decoherence and dephasing across the ensemble.  

Our atomic qubits are embedded within a larger ground hyperfine manifold, and can be coupled to states outside the logical space either by the microwave driving field (see below) or by optical pumping due to scattering of photons from the probe and optical lattice light fields.  It is straightforward to solve the master equation for a microwave driven atom in the presence of optical pumping; we have done so for parameters typical of our experiment and have found that the qubit pseudospin can be approximated to an excellent degree as a lossy $2$-level system.  This allows us to replace the numerically intensive master equation model with a much simpler effective model based on the Torrey solutions to the familiar Bloch equations \protect\cite{Torrey1949} \protect\cite{Allen1987}.  In our experiments the microwave resonant Rabi frequency is always much larger than the inverse of the longitudinal ($T_1$) and transverse ($T'_2$) homogeneous lifetimes, so that $\chi\gg1/T'_2-1/T_1$ (the strong driving limit of the Torrey solutions).  The detuning of the microwave field relative to the qubit transition frequency is also relatively small, $\Delta\sim10^{-1}$.  With the initial condition $\vert\psi_i\rangle=\vert0\rangle$, the Torrey solution for the expectation value of the $\mathbf{3}$-component of the qubit pseudospin (the inversion) is then
\begin{equation}
w(t) = -\frac{\Delta^2}{\Omega^2}\exp^{-\frac{2}{3}\gamma_1t}
+\frac{\chi^2}{\Omega^2}\cos(\Omega t)\exp^{-\gamma_1 t}+2\frac{\Delta^2}{\Omega^2}
\label{eq:inversion}
\end{equation}
where $\Omega=\sqrt{\chi^2+\Delta^2}$ and $\gamma_1=3/(4T_1)$, and where we have set $T'_2=2T_1$ as appropriate for a closed two-level system subject only to optical pumping.  If the total population of the two-level system decays at a rate $\gamma_2$, then the logical-$\vert1\rangle$ population oscillates in time as
\begin{align}
\Pi_1(t) &=\frac{1}{2}[w(t)+1]e^{-\gamma_2 t}\nonumber \\ &=\frac{1}{2}(1-2\frac{\Delta^2}{\Omega^2})e^{-\gamma_2 t}-\frac{1}{2}\frac{\chi^2}{\Omega^2}\cos(\Omega t)e^{-(\gamma_1+\gamma_2) t}\nonumber \\&\qquad+\frac{1}{2}\frac{\Delta^2}{\Omega^2}e^{-(\frac{2}{3}\gamma_1+\gamma_2) t}\label{eq:populationa}\\&\approx\frac{1}{2}e^{-\gamma_2 t}-\frac{1}{2}\frac{\chi^2}{\Omega^2}\cos(\Omega t)e^{-(\gamma_1+\gamma_2)t},
\label{eq:populationb}
\end{align}
where the last approximation introduces errors of order $10^{-3}$ for the parameters typical of our experiments.

Inhomogeneous effects occur due to variations in the microwave resonant Rabi frequency and/or detuning.  We therefore define an ensemble averaged population
\begin{equation}
\bar{\Pi}_1(t)=\int_{-\infty}^\infty d\chi\int_{-\infty}^\infty d\Delta \, P(\chi,\Delta)\Pi_1(t),
\label{eq:average}
\end{equation}
where the distribution $P(\chi,\Delta)$ is Gaussian, with mean values $\chi_0$, $\Delta_0$ and standard deviations $\delta\chi$, $\delta\Delta$ for the resonant Rabi frequency and detuning, respectively.  In the end, our expected polarimetry signal is of the form
\begin{equation}
S(t)=S_1\bar{\Pi}_1(t)+S_0
\label{eq:signal}
\end{equation}
where the signal amplitude $S_1$ and offset $S_0$ depend on probe power and balancing of the polarimeter.  The function $S(t)$ is used to fit our experimental data sets, with free parameters $\chi_0$, $\Delta_0$, $\delta\chi$, $\delta\Delta$, $S_1$, and $S_0$.  The homogeneous decay rates $\gamma_1$ and $\gamma_2$ are not included as free parameters in the fit, but instead determined independently in a separate spin-echo experiment as described in Sec. IV.

\begin{figure}
[t]\resizebox{8.75cm}{!}
{\includegraphics{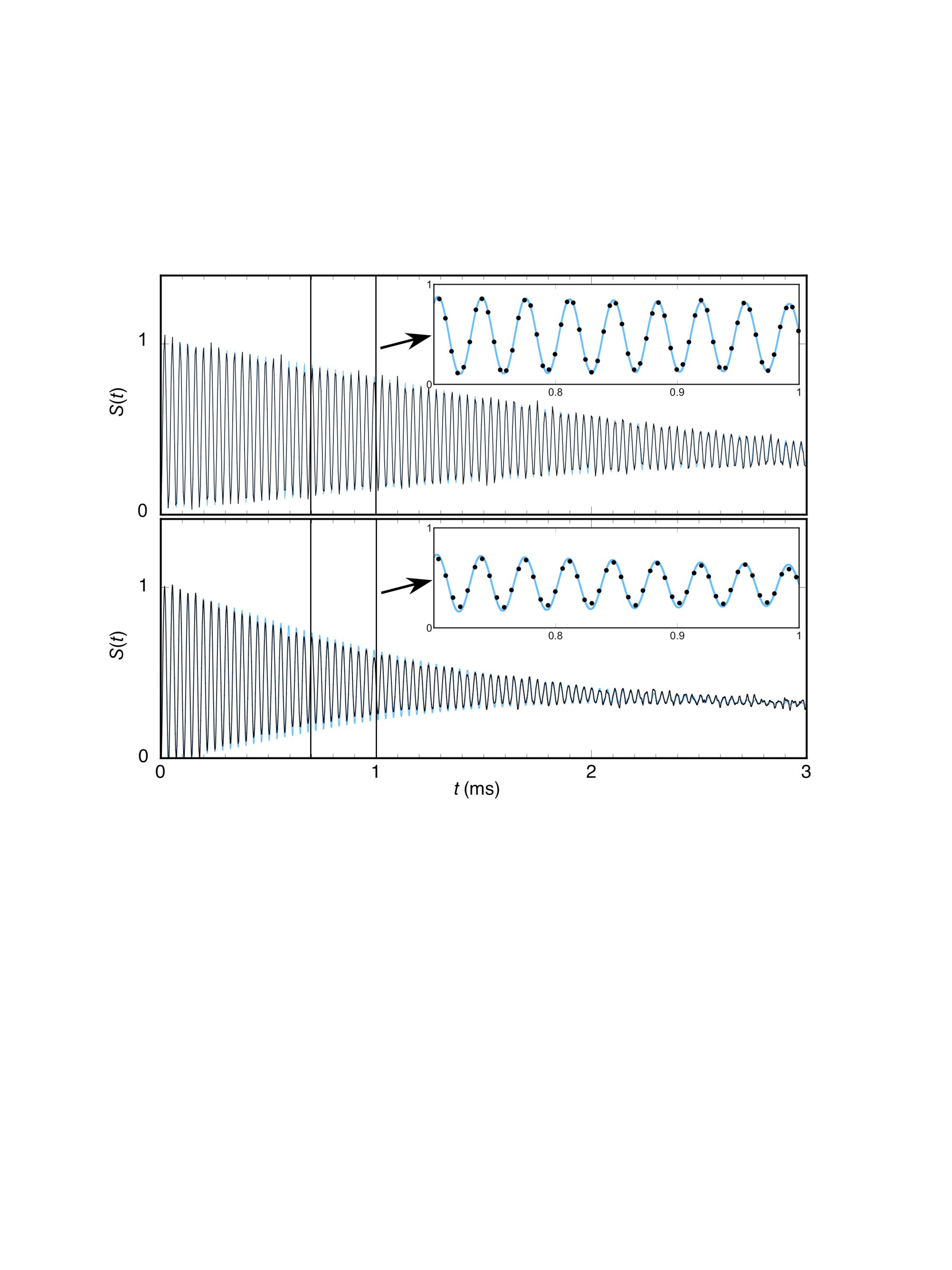}}
\caption{\label{fig:rabiosc}(Color online) Polarimetry signal $S(t)$ measured during qubit Rabi oscillation, for atoms in free fall (top) and atoms trapped in the optical lattice (bottom).  The dark line is experimental data, and the underlying light line is the result of a fit of the form given in Eq.~\eqref{eq:signal}.  Inserts show a close-up of data points (filled circles) and fit for a representative time interval.}
\end{figure}

Figure \ref{fig:rabiosc} shows typical examples of our polarimetry signal for atoms in free fall, and for atoms trapped in the optical lattice.   Two features are immediately apparent in both data sets: the oscillation amplitude decays in a few milliseconds, corresponding to several tens of Rabi periods, and the mean of the oscillation decays on a much slower timescale.  This is characteristic of situations where dephasing plays a significant role.  The figure also shows fits of the form $S(t)$ to each signal.  In the case of free atoms such fits imply a microwave irradiance inhomogeneity across the atomic sample of $\delta\chi/\chi_0=0.3\%$, and a detuning inhomogeneity, $\delta\Delta/\chi_0=3.3\%$, presumably caused by spatial variations in our bias magnetic field.  For atoms trapped in the optical lattice we find $\delta\chi/\chi_0=0.3\%$ and $\delta\Delta/\chi_0=7.3\%$, with the increase in $\delta\Delta$ reflecting the additional inhomogeneous broadening from spatial variations in the lattice light shift.  These values for free and trapped atoms represent averages over a number of measurements taken under nominally identical conditions but at widely different times.  

Based on the parameters $\chi_0$, $\Delta_0$, $\delta\chi$, and $\delta\Delta$ extracted from high-quality fits to Rabi oscillation data, and on the homogeneous decay rates $\gamma_1$ and $\gamma_2$ determined from spin echoes, we can accurately model the ensemble averaged population $\bar{\Pi}_1(t)$ of the logical-$\vert1\rangle$ state.  This in turn allows us to obtain a good estimate of the ensemble averaged fidelity $\mathscr{F}_{\pi}$ for a perfectly timed $\pi$-gate, defined as a square microwave pulse of duration $\tau=\pi/\chi_0$ applied across the entire ensemble.   Note that in doing so, the relevant decay rates are those obtained in the presence of the optical lattice only, since the probe beam would not be present in experiments aimed at quantum information processing.  We have, in this fashion, found a $\pi$-gate fidelity $\mathscr{F}_{\pi}=0.990(3)$ for atoms in the optical lattice, averaged over $16$ independent measurements taken over a $3$ month period.  This value is a measure of the limitations imposed by working with qubits in spatially inhomogeneous lattice and microwave fields.

In practice, the fidelity of a gate operation is further reduced if we cannot preset the mean Rabi frequency $\chi_0$ and detuning $\Delta_0$ with sufficient accuracy.  From our data set we find statistical fluctuations of $1\%$ rms in the fitted value for the mean Rabi frequency $\chi_0$, relative to our target value of $27.78kHz$ which corresponds to an $18.0\mu s$ $\pi$-gate time.  In the same data set the rms fluctuations in the mean detuning were $\Delta_0/\chi_0\sim4.8\%$.  We can use the fit parameters for each individual measurement to calculate $\bar{\Pi}_1(18.0\mu s)$, which is the ensemble-averaged fidelity at the time intended to produce a $\pi$-gate.  Averaged over our $16$ independent measurements, this reduces the fidelity by a statistically insignificant $1\times10^{-4}$.

In our discussion so far, we have ignored the possibility of microwave coupling to states outside the qubit logical space.  For our encoding such coupling occurs only between the $\vert F=3,m_F=3\rangle$ logical-$\vert0\rangle$  state and the states $\vert F=4,m_F=2,3\rangle$.  These transitions are Zeeman shifted out of resonance by the applied bias magnetic field, and further suppressed by maximizing the $\sigma_+$-polarized component of the microwave field.  For a quantitative estimate, it is  straightforward to tune our microwave frequency to be resonant with these transitions and to determine the resonant Rabi frequencies in each case.  For the data sets discussed here, we thus found Rabi frequencies of $3kHz$ for both transitions, and detunings of $130kHz$ and $260kHz$ respectively.  During a $\pi$-gate on the qubit this translates into a combined population of $7\times10^{-4}$ in the two states, and thus a reduction in the $\pi$-gate fidelity of less than $1\times10^{-3}$.  This is well below the statistical uncertainty of our present experiment, but if necessary the population leakage can be further suppressed with a stronger bias magnetic field.

\section{IV. Rotary Echoes and Coherence Times}
\begin{figure}
[t]\resizebox{8.75cm}{!}
{\includegraphics{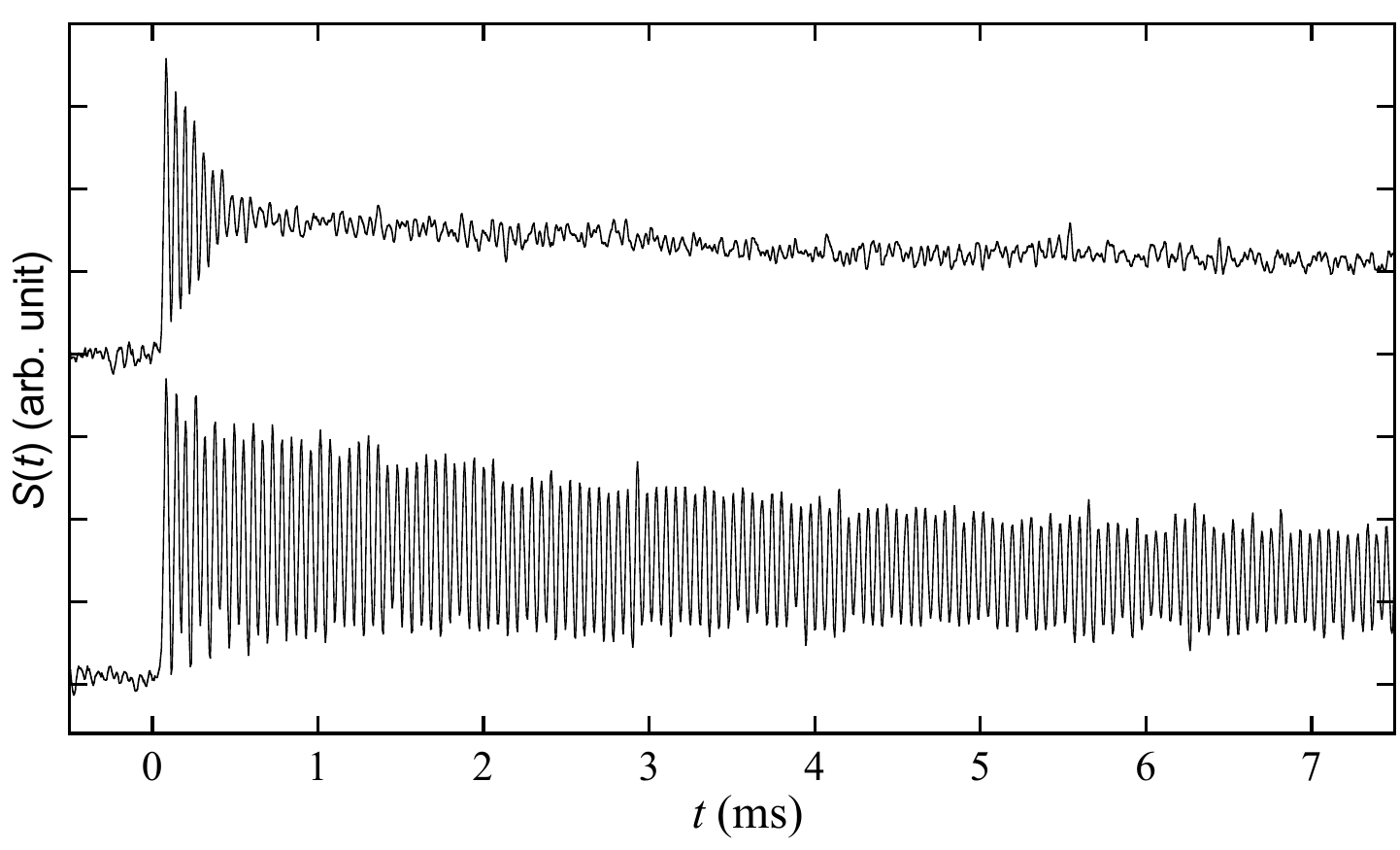}}
\caption{\label{fig:rabirotary}Polarimetry signal $S(t)$ during qubit Rabi oscillation (top) and a rotary-$2\pi$  spin-echo sequence.  An artificially large inhomogeneity in the microwave detuning $\Delta$ has been added to demonstrate the robustness of the rotary-$2\pi$  echo.}
\end{figure}

As the preceding discussion makes clear, the decay of qubit Rabi oscillations in our optical lattice is dominated by dephasing due to inhomogeneities in the microwave irradiance and light shift across the ensemble.   In this situation the underlying coherence times cannot be reliably determined by fitting experimental measurements of Rabi oscillations with the model given in \eqref{eq:populationa}.  To circumvent this problem we use a technique originally developed by the NMR community, rotary echoes \protect\cite{chuangrevmod}, to suppress dephasing and obtain independent estimates for the rates $\gamma_1$ and $\gamma_2$.  The simplest rotary echo sequence consists of repeated application of the composite rotation $R_\mathbf{u}(\theta)R_{\mathbf{u}'}(\theta)$, where the axes of rotation are controlled through the phase $\varphi$ of the driving field.  For given Rabi frequency and detuning we have 
\begin{equation}
\mathbf{u}=\frac{\vert\chi\vert}{\Omega}[\cos(\varphi)\mathbf{1}+\sin(\varphi)\mathbf{2}]+\frac{\Delta}{\Omega}\mathbf{3}
\label{eq:axis}
\end{equation}
In the near resonance case, $\Delta\ll0$, a phase change $\varphi\rightarrow\varphi+\pi$ will nearly invert the axis of rotation, $\mathbf{u}'\approx\mathbf{u}$, and the composite rotation remains close to the identity. Thus, if the qubit is initialized in logical-$\vert0\rangle$ and $\theta$ is an integer multiple of $\pi$, the rotary echo sequence will produce a polarimetry signal that is indistinguishable from standard Rabi oscillations with greatly suppressed inhomogeneity in $\chi$.  In general, rotary echoes dephase as rapidly due to inhomogeneity in $\Delta$ as do Rabi oscillations.  Remarkably, however, the special case $\theta=2\pi$ is robust against variations in $\chi$ \emph{and} $\Delta$, and a rotary-$2\pi$ echo sequence looks like Rabi oscillations with greatly suppressed inhomogeneity in both.  This is easy to verify through numerical simulation of rotary-$2\pi$  echoes, which for inhomogeneities typical of our experiment show no sign of dephasing on timescales far beyond the longest coherence times that we have observed in the laboratory.   Figure \ref{fig:rabirotary} is an experimental illustration of this robustness, comparing polarimetry signals from Rabi oscillations and a rotary-$2\pi$  echo in the presence of an artificially increased inhomogeneity in $\Delta$.

\begin{figure}
[t]\resizebox{8.75cm}{!}
{\includegraphics{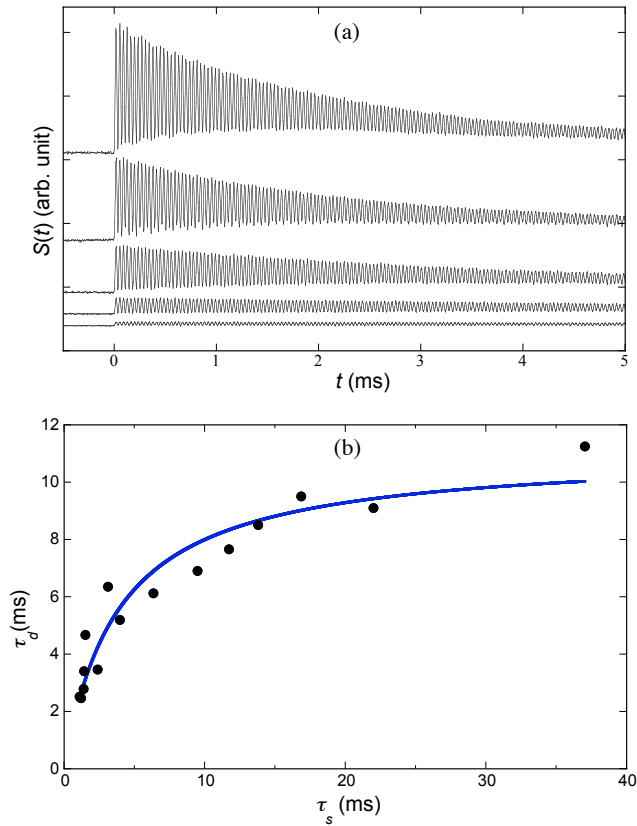}}
\caption{\label{fig:decaytimes}(a) Polarimetry signals $S(t)$ measured during rotary-$2\pi$ spin-echo sequences.  The probe power is reduced by a factor of $\sim40$ from top to bottom, to demonstrate the longer coherence times and lower signal levels available at lower rates of probe photon scattering.  (b) Coherence time $\tau_d$ versus probe photon scattering time $\tau_s$.}
\end{figure}

The analysis of qubit Rabi oscillations in Sec. III requires that we determine the rates $\gamma_1$ and $\gamma_2$ in the presence of a probe beam of identical intensity and frequency.  This is easily achieved by recording the measurement from Rabi oscillation and a rotary-$2\pi$ echo in quick succession, and fitting the latter by setting $\delta\chi=\delta\Delta=0$ in our model for Rabi oscillation.  To estimate gate fidelities we also need to determine these decay rates in the limit where photon scattering from the probe becomes negligible.  Figure \ref{fig:decaytimes}(a) shows rotary-$2\pi$  echo signals for a few probe powers, clearly illustrating the decrease in signal strength and increase in coherence time as the probe scattering rate is reduced.  Figure \ref{fig:decaytimes}(b) shows the time constant $\tau_d$ for the decay of the oscillating part of the echo signal, as a function of the scattering time (the inverse probe scattering rate) $\tau_s=1/\gamma_s$.  As one would expect, $\tau_d$ is roughly proportional to $\tau_s$ for short scattering times, and saturates for long scattering times.  Also shown is a fit of the form $\tau_d=1/(a\gamma_s+b)$, which for this data set gives an asymptotic value $\tau_d\approx10.6ms$ that presumably reflects the coherence time for qubits in the optical lattice.  Note that this is only a representative data set; over time and for nominally identical conditions we have observed asymptotes in the range $5.5ms-15ms$.  For the data analysis in Sec. III we therefore used the most conservative estimate, $\tau_d=5.5ms$.  For technical reasons (low-frequency noise and weak polarimetry signals at low $\gamma_s$) it is more challenging to get an accurate estimate for the decay of the constant part of the echo signal.  We have generally found this decay time to be $\sim2\times\tau_d$, and therefore $\gamma_1\approx \gamma_2\approx1/(2\tau_d)$ according to Eq. \eqref{eq:populationb}, and we have assumed this throughout our data analysis.  This observation is also consistent with a full master equation model of the dynamics of the entire $6S_{1/2}$ ground manifold in the presence of optical pumping and a microwave drive.

It is instructive to consider what physical mechanisms might limit the qubit coherence time. First, for our encoding the qubit transition frequency is sensitive to temporal fluctuations in the lattice depth and in the background magnetic field.  As long as the power spectrum of these fluctuations lies well below the Rabi frequency, driving continuous Rabi oscillations or rotary echoes is equivalent to applying a sequence of refocusing pulses, and one expects to see longer coherence times than in, e. g., a Ramsey interrogation \protect\cite{Viola1999}.  Our experiment does indeed show such an increase.  Second, qubits decohere due to optical pumping caused by the scattering of photons from the optical lattice.  For our relatively deep and near-resonance optical lattice we calculate a scattering time of $\sim100ms$, which is too large to account for the $\sim10ms$ coherence times seen in our experiment.  Further evidence of the minimal role played by lattice photon scattering comes from the fact that the coherence time remains effectively constant when the scattering rate is changed by as much as a factor of two.  Finally, we note that phase noise in the microwave source will cause the axis of rotation to fluctuate and thus degrade the rotary echo.  If the phase noise spectrum is flat we expect the echo signal to degrade after a roughly constant number of echo pulses, resulting in longer coherence times when the microwave Rabi frequency is reduced.  The qualitative behavior of our experiment suggests that this may be an important contributing factor in limiting the coherence time, but our attempts to directly measure the phase noise and model its influence on the qubit dynamics have so far been unsuccessful.

\section{V. Composite Pulses}
Composite rotations are a powerful tool to overcome the effect of inhomogeneous errors, not just in spin-echo and refocusing sequences, but also in the implementation of quantum gates.   Cummins \protect\cite{Cummins2003} describes three families of universal pulse sequences which are designed to confer robustness in different scenarios.  The term \emph{universal} here refers to a pulse sequence that can implement any desired unitary transformation, i. e. any rotation about a given axis by a given angle, with a similar degree of error compensation for all initial states.  This contrasts with pulse sequences designed to be robust only for the transfer of a specific input state to a specific output state, e. g. the familiar 3-pulse refocusing sequence $R_\mathbf{1}(\pi/2)R_\mathbf{2}(\pi)R_\mathbf{1}(\pi/2)$. 

\begin{figure}
[t]\resizebox{8.75cm}{!}
{\includegraphics{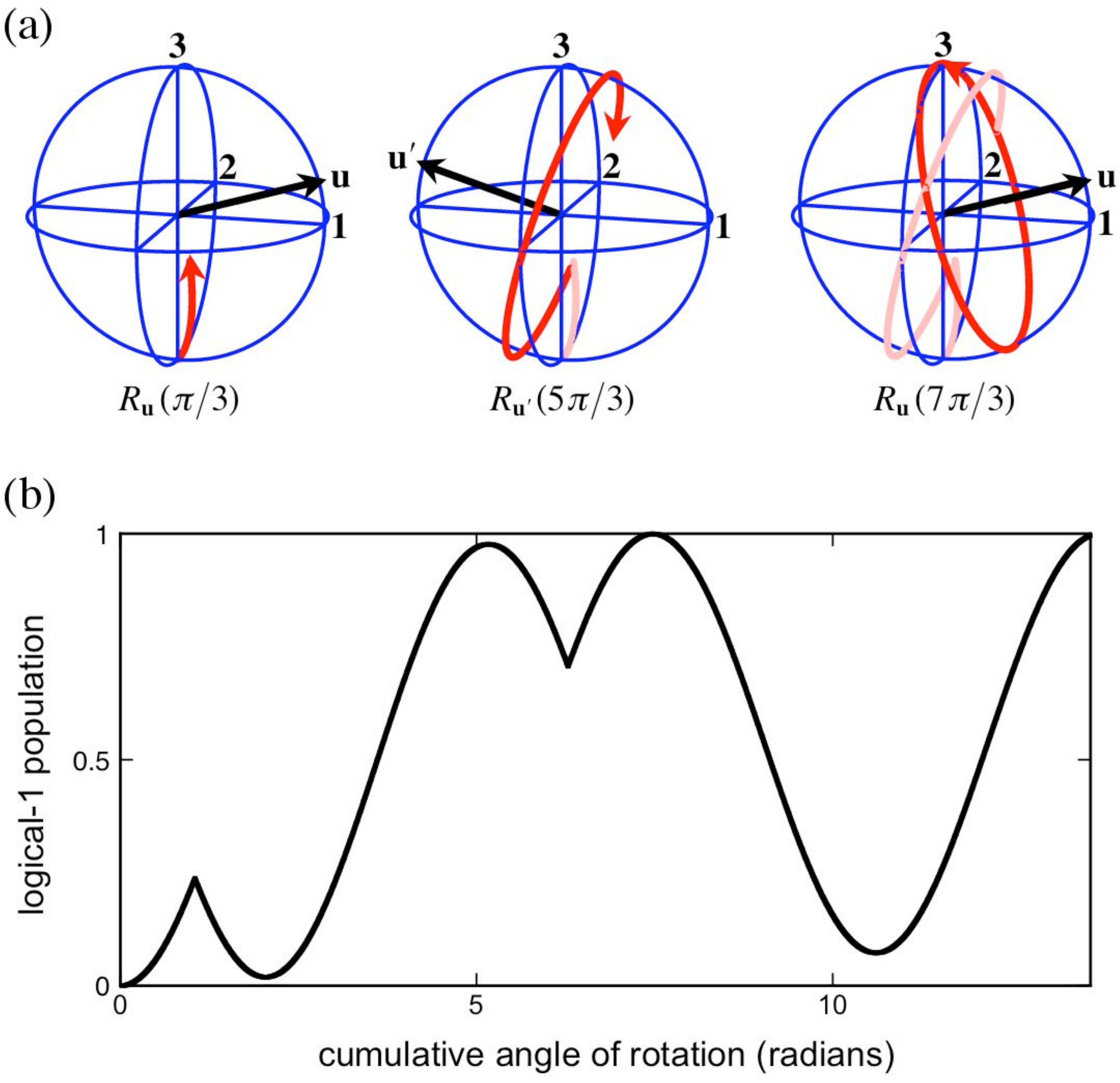}}
\caption{\label{fig:corpsebloch}(Color online) (a) Trajectories on the Bloch sphere during the three parts of a composite CORPSE $\pi$-pulse, in the presence of a detuning $\Delta=0.3\chi$.  (b) Evolution of the logical-$\vert1\rangle$ population during the pulse.}
\end{figure}

For our system the dominant factor limiting gate fidelity is the variation of effective microwave detuning across the qubit ensemble.  The so-called CORPSE (Compensation for Off-Resonance with a Pulse Sequence) family is a three-pulse sequence that confers robustness against these detuning errors.  If, for example, our goal is to perform a perfect $\pi$ rotation around the $\mathbf{1}$-axis, then we can approximate this through the CORPSE sequence
\begin{equation}
R_\mathbf{u}(\pi/3)R_{\mathbf{u}'}(5\pi/3)R_\mathbf{u}(7\pi/3)\approx R_\mathbf{1}(\pi),
\label{eq:corpse}
\end{equation}
where $\mathbf{u}=(\vert\chi\vert/\Omega)\mathbf{1}+(\Delta/\Omega)\mathbf{3}$, $\mathbf{u}'=-(\vert\chi\vert/\Omega)\mathbf{1}+(\Delta/\Omega)\mathbf{3}$.  Figure \ref{fig:corpsebloch} shows the corresponding trajectory on the Bloch sphere, as well as the evolution of the logical-$\vert1\rangle$ population during the pulse.  To leading order in the detuning, $f=\Delta/\chi_0\ll1$, the fidelity of a single square pulse is $\mathscr{F}\approx1-f^2/2$.  By comparison, a CORPSE $\Theta$-pulse achieves $\mathscr{F}\approx1-\alpha f^4$, where $\alpha=6.5\times10^{-3}$ for $\Theta=\pi$, at the cost of increased sensitivity to errors in the rotation angle.  Similarly, the SCROFULOUS (Short Composite Rotation For Undoing Length Over and Under Shoot) family is a three-pulse sequence that confers robustness against errors in the rotation angle at the cost of increased sensitivity to detuning errors, and the BB1 (BroadBand Number 1) family is a five-pulse sequence that compensates angle errors to sixth order at little or no cost in sensitivity to detuning errors.  We refer the reader to the literature \protect\cite{Cummins2003}\protect\cite{chuangrevmod} for further discussion of these pulse families, including how to design composite rotations by angles other than $\pi$.  We note also that new numerical techniques enable the brute-force optimization of much more complicated time dependent phase variations, which can improve the robustness to specific combinations of amplitude and detuning errors, or achieve better robustness in the presence of decoherence \protect\cite{Khaneja2004}.

\begin{figure*}
[t]\resizebox{17.5cm}{!}
{\includegraphics{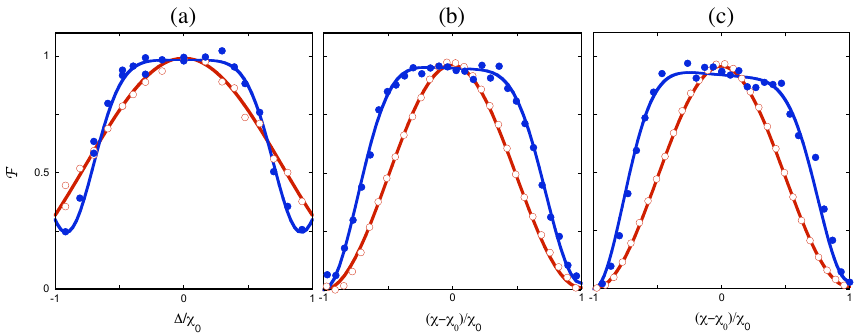}}
\caption{\label{fig:composite} (Color online) Fidelity of a plain and three types of composite $\pi$-pulses, as function of deliberately applied errors in either the average detuning $\Delta_0$ or average resonant Rabi frequency $\chi_0$.  The pulse sequences are (a) CORPSE, (b) SCROFULOUS, and (c) BB1.  Filled and open circles are experimentally measured fidelities for the composite pulses and plain pulses,  respectively.  Solid lines are theoretical fidelities found by solving the optical Bloch equations, using independently determined inhomogeneities and coherence times.}
\end{figure*}

It is straightforward to check the performance of composite pulses in our system by preparing our atoms in logical-$\vert0\rangle$, apply the appropriate pulse sequence, and measure the logical-$\vert1\rangle$ population.  Figure \ref{fig:composite} shows the fidelity of plain pulse, CORPSE, SCROFULOUS and BB1 $\pi$-gates for large, intentionally applied errors.  Each experimental data set closely follows the predictions of our theoretical model, and illustrates how a given pulse family significantly extends the range of detuning or angle error over which one can achieve good fidelity.  It is worth noting, however, that for SCROFULOUS and BB1 pulses the peak fidelity falls somewhat below that of a plain pulse.  This reduction is due to a combination of factors: these composite pulses protect against angle errors, which are insignificant when the intentional error is near zero, the SCROFULOUS pulse increases sensitivity to the intrinsic detuning spread in our experiment, and both take longer to execute (for a given Rabi frequency) and therefore subject qubits to additional decoherence during the gate operation.  In our case, the gain from increased robustness is more than offset by the loss from decoherence, and the overall fidelity goes down.  The tradeoff is more favorable for CORPSE pulses, where there is some gain from increased robustness to the intrinsic detuning errors that dominate in our experiment.

\begin{figure}
[t]\resizebox{8.75cm}{!}
{\includegraphics{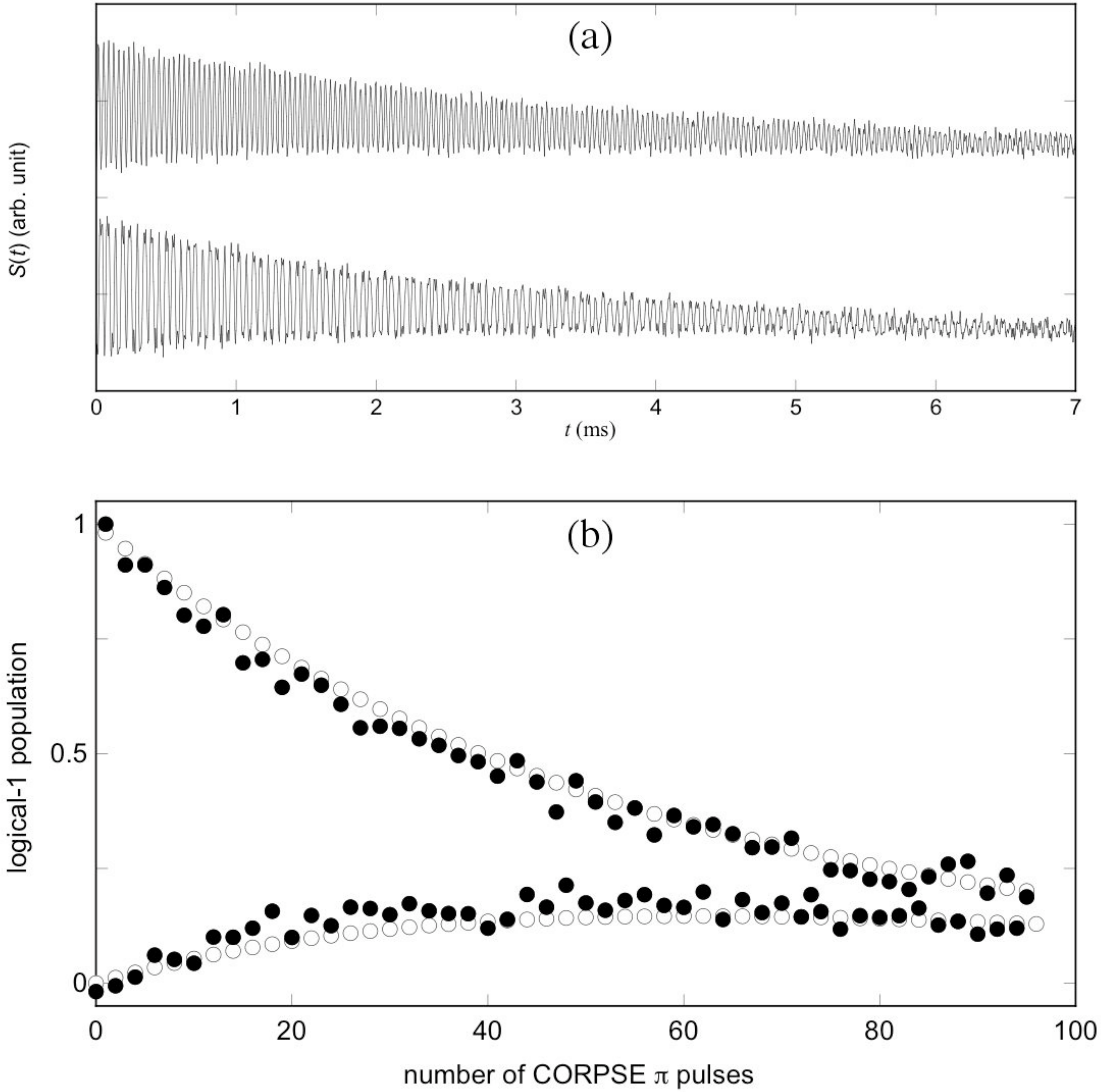}}
\caption{\label{fig:rotarycorpse}(a) Polarimetry signals $S(t)$ measured during a rotary-$2\pi$ spin-echo sequence (top) and a sequence of CORPSE $\pi$-pulses.  The nearly identical decay times demonstrate that the CORSE pulse successfully compensates for detuning errors, with little or no cost incurred from the increased sensitivity to angle errors.  (b)  Population of the logical-$\vert1\rangle$ state after an integral number of CORPSE $\pi$-pulses.  Solid circles are experimental data extracted from the signal in (a).  Open circles are theoretical values found by solving the optical Bloch equations, using independently determined inhomogeneities and coherence times.}
\end{figure}

High fidelity CORPSE $\pi$-pulses can be evaluated in much the same way as plain $\pi$-pulses: by successive application and measurement of the logical-$\vert0\rangle$ and -$\vert1\rangle$ populations.  Figure \ref{fig:rotarycorpse}(a) compares polarimetry signals for a rotary-$2\pi$  echo, and for successive CORPSE pulses.  Note the similar decay times for the two pulse sequences, as expected when detuning inhomogeneity dominates.  In principle the CORPSE data can be analyzed using an approach analogous to Sec. III, but in practice it is difficult to fit this rather complicated polarimetry signal.  Instead, we read off the logical-$\vert0\rangle$  and -$\vert1\rangle$  populations at the beginning and end of each CORPSE pulse, and compare these to the predictions of a full model based on the optical Bloch equations, using the inhomogeneities and relaxation rates determined from Rabi oscillations and rotary-$2\pi$ echoes performed under identical conditions.  Figure \ref{fig:rotarycorpse}(b) shows a typical data set for which these values agree quite closely.  This provides confidence that our model correctly describes the performance of CORPSE pulses in our experiment, and allows us to estimate a CORPSE fidelity $\mathscr{F}_{CORPSE}=0.992$.  Since the inhomogeneities and relaxation rates for this data set are typical of our larger data set, we conclude that CORPSE pulses will yield only marginal improvement for our conditions.  However, in situations where inhomogeneities are larger or coherence times longer, it appears likely that CORPSE or other composite pulse sequences can significantly improve the fidelities of single-qubit rotations.

\section{VI. Summary and Outlook}
We have encoded qubits in the ground state manifold of neutral Cs atoms trapped in an optical lattice, and implemented accurate single-qubit $\pi$-gates simultaneously across the entire ensemble by driving the atoms with microwaves from a phase agile, programmable source.  The fidelity of these gates was found to depend primarily on spatial variations in the qubit transition frequency which is Zeeman shifted by applied and background magnetic fields, and light-shifted by the presence of the optical lattice.  Spatial variations in the microwave Rabi frequency and decoherence from scattering of lattice light played a lesser role in limiting the fidelity of simple rotations.  Accurate estimates of the $\pi$-gate fidelity were obtained from quantitative modeling of Rabi oscillation data.  When averaged over many data sets taken over a 3 month period, this yielded a value $\mathscr{F}_{\pi}=0.990(3)$.  Independent estimates of coherence times were obtained from rotary-$2\pi$  spin echo sequences that effectively compensates for inhomogeneities.  Finally, we implemented three types of composite pulses and tested their ability to compensate for large, deliberately applied errors in Rabi frequency or detuning.  The CORPSE sequence provides increased robustness against detuning errors, and is potentially the most useful in an atom/lattice system where the plain pulse fidelity is limited by detuning inhomogeneity across the ensemble.  In our experiment the inhomogeneities and coherence times are such that the advantage from increased robustness is outweighed by additional decoherence during the longer CORPSE sequence, and we were not able to achieve any significant gain in fidelity.  However, this situation may not be representative, and optical lattice experiments elsewhere could offer tradeoffs that favor the use of composite pulse techniques.

The single-qubit gate fidelities demonstrated here are probably good enough that they will not limit neutral atom quantum information processing in the near future.  Nevertheless, single-qubit control being the simplest task on this particular platform, it is important to consider how one might improve the performance significantly.  The most straightforward approach is probably to increase the microwave Rabi frequency, by using a more powerful microwave source and/or moving the gain horn closer to the atomic sample.  Increasing the Rabi frequency ten-fold while keeping other parameters unchanged should reduce the single-pulse $\pi$-gate time to $\sim5.7\mu s$, and improve fidelity by nearly an order of magnitude, to $\mathscr{F}=0.9987$.  Next, one might attempt to decrease inhomogeneities in Rabi frequency and detuning, most likely by working with a more compact sample of atoms, or to increase the coherence time enough for the robustness versus decoherence tradeoff to favor the use of composite pulses.  Much longer coherence times should be readily achievable if one chooses a magnetic field insensitive qubit encoding, employs far detuned optical traps with lower scattering rates, and improves the phase noise of the microwave source.  Furthermore, the use of relaxation optimized pulse sequences has the potential to significantly reduce the decoherence incurred for a desired level of robustness \protect\cite{Khaneja2004}.

A key challenge for quantum information processing in optical lattices is to develop methods to address and manipulate individual atomic qubits.  In large period optical lattices such addressable gates might be implemented with optical Raman transitions and focused laser beams \protect\cite{Nelson2007}, in a manner similar to that used in ion traps \protect\cite{Nagerl1999}.  Alternatively, one might use magnetic field gradients \protect\cite{Schrader2004}, or the light shift in a tightly focused laser beam \protect\cite{Zhang2006}, to map qubit position into a shift in transition frequency, and use microwaves to selectively manipulate an individual qubit or a group of qubits.  In practice it will be challenging to accurately position a focused laser beam relative to an optical lattice, leading to significant uncertainty in the frequency shift associated with a given qubit position.  In that case it becomes essential to use advanced composite pulse techniques that ensure high gate fidelity for the targeted qubit, and simultaneous suppression of spurious rotations of the neighboring qubits.  The CORPSE response of Fig. \ref{fig:composite}(a) serves as a rough illustration of the desired behavior: it offers high fidelity for a limited range of detunings, and steep roll-off outside the acceptance bandwidth.  However, the CORPSE sequence is not optimized to suppress qubit response everywhere outside the acceptance bandwidth, and is therefore unsuitable for robust addressing.  Preliminary work suggests that numerical optimization of more complex pulse phase modulation can be used to design composite pulses with nearly ideal response \protect\cite{Deutsch2008}.  We hope to explore the use of numerically optimized composite pulses in future experiments aimed at higher fidelity and addressable single-qubit gates.

The authors thank Ivan H. Deutsch and Navin Khaneja for helpful discussions.  This work was funded by IARPA Grant No. DAAD19-13-R-0011, and NSF Grant No. PHY-0555673.



\begin{thebibliography}{21}
\expandafter\ifx\csname natexlab\endcsname\relax\def\natexlab#1{#1}\fi
\expandafter\ifx\csname bibnamefont\endcsname\relax
  \def\bibnamefont#1{#1}\fi
\expandafter\ifx\csname bibfnamefont\endcsname\relax
  \def\bibfnamefont#1{#1}\fi
\expandafter\ifx\csname citenamefont\endcsname\relax
  \def\citenamefont#1{#1}\fi
\expandafter\ifx\csname url\endcsname\relax
  \def\url#1{\texttt{#1}}\fi
\expandafter\ifx\csname urlprefix\endcsname\relax\def\urlprefix{URL }\fi
\providecommand{\bibinfo}[2]{#2}
\providecommand{\eprint}[2][]{\url{#2}}

\bibitem[{\citenamefont{{G. K. Brennen, C. M. Caves, P. S. Jessen, and I. H. Deutsch}}(1999)}]{Brennen1999}
\bibinfo{author}{\bibnamefont{{G. K. Brennen, C. M. Caves, P. S. Jessen and I. H. Deutsch}}},
\bibinfo{journal}{Phys.\ Rev.\ Lett.} \textbf{\bibinfo{volume}{82}},
\bibinfo{pages}{1060} (\bibinfo{year}{1999}).
  
\bibitem[{\citenamefont{{D. Jaksch, H.-J. Briegel, J. I. Cirac, C. W. Gardiner, and P. Zoller}}(1999)}]{Jaksch1999}
\bibinfo{author}{\bibnamefont{{D. Jaksch, H.-J. Briegel, J. I. Cirac, C. W. Gardiner, and P. Zoller}}},
\bibinfo{journal}{Phys.\ Rev.\ Lett.} \textbf{\bibinfo{volume}{82}},
\bibinfo{pages}{1975} (\bibinfo{year}{1999}).
  
\bibitem[{\citenamefont{{O. Mandel, M. Greiner, A. Widera, T. Rom, T.W. Hansch, and I. Bloch}}(1999)}]{Mandel2003}
\bibinfo{author}{\bibnamefont{{O. Mandel, M. Greiner, A. Widera, T. Rom, T.W. Hansch, and I. Bloch}}},
\bibinfo{journal}{Nature} \textbf{\bibinfo{volume}{425}},
\bibinfo{pages}{937} (\bibinfo{year}{2003}).
  
\bibitem[{\citenamefont{{M. Anderlini, P. J. Lee, B. L. Brown, J. Sebby-Strabley, W. D. Phillips, and J. V. Porto}}(1999)}]{Anderlini2007}
\bibinfo{author}{\bibnamefont{{M. Anderlini, P. J. Lee, B. L. Brown, J. Sebby-Strabley, W. D. Phillips, and J. V. Porto}}},
\bibinfo{journal}{Nature} \textbf{\bibinfo{volume}{448}},
\bibinfo{pages}{452} (\bibinfo{year}{2007}).

\bibitem[{\citenamefont{{D. Jaksch, J. I. Cirac, P. Zoller, S. L. Rolston, R. Cote, and M. D. Lukin}}(2000)}]{Jaksch2000}
\bibinfo{author}{\bibnamefont{{D. Jaksch, J. I. Cirac, P. Zoller, S. L. Rolston, R. Cote, and M. D. Lukin}}},
\bibinfo{journal}{Phys.\ Rev.\ Lett.} \textbf{\bibinfo{volume}{85}},
\bibinfo{pages}{2208} (\bibinfo{year}{2000}).

\bibitem[{\citenamefont{{M. D. Lukin, M. Fleischhauer, R. Cote, L. M. Duan, D. Jaksch, J. I. Cirac, and P. Zoller}}(2001)}]{Lukin2001}
\bibinfo{author}{\bibnamefont{{M. D. Lukin, M. Fleischhauer, R. Cote, L. M. Duan, D. Jaksch, J. I. Cirac, and P. Zoller}}},
\bibinfo{journal}{Phys.\ Rev.\ Lett.} \textbf{\bibinfo{volume}{87}},
\bibinfo{pages}{037901} (\bibinfo{year}{2001}).

\bibitem[{\citenamefont{{E. Urban, T. A. Johnson, T. Henage, L. Isenhower, D. D. Yavuz, T. G. Walker, and M. Saffman}}(2008)}]{Urban2008}
\bibinfo{author}{\bibnamefont{{E. Urban, T. A. Johnson, T. Henage, L. Isenhower, D. D. Yavuz, T. G. Walker, and M. Saffman}}},
\bibinfo{journal}{arXiv:0805.0758v1 [quant-ph]}, (\bibinfo{year}{2008}).

\bibitem[{\citenamefont{{A. Gaetan, Y. Miroshnychenko, T. Wilk, A. Chotia, M. Viteau, D. Comparat, P. Pillet, A. Browaeys, and P. Grangier}}(2008)}]{Gaetan2008}
\bibinfo{author}{\bibnamefont{{A. Ga\"etan, Y. Miroshnychenko, T. Wilk, A. Chotia, M. Viteau, D. Comparat, P. Pillet, A. Browaeys, and P. Grangier}}},
\bibinfo{journal}{arXiv:0810.2960v1 [quant-ph]}, (\bibinfo{year}{2008}).

\bibitem[{\citenamefont{{H. K. Cummins, G. Llewellyn, and J. A. Jones}}(2003)}]{Cummins2003}
\bibinfo{author}{\bibnamefont{{H. K. Cummins, G. Llewellyn, and J. A. Jones}}},
\bibinfo{journal}{Phys.\ Rev.\ A.} \textbf{\bibinfo{volume}{67}},
\bibinfo{pages}{042308} (\bibinfo{year}{2003}).

\bibitem[{\citenamefont{{L.M.K. Vandersypen and I.L. Chuang}}(2004)}]{chuangrevmod}
\bibinfo{author}{\bibnamefont{{L.M.K. Vandersypen and I.L. Chuang}}},
\bibinfo{journal}{Rev.\ Mod.\ Phys.} \textbf{\bibinfo{volume}{76}},
\bibinfo{pages}{1037} (\bibinfo{year}{2004}).
  
\bibitem[{\citenamefont{{S. Gulde, M. Riebe, G. P. T. Lancaster, C. Becher, J. Eschner, H. Haffner, F. Schmidt-Kaler, I. L. Chuang, and R. Blatt}}(2003)}]{Gulde2003}
\bibinfo{author}{\bibnamefont{{S. Gulde, M. Riebe, G. P. T. Lancaster, C. Becher, J. Eschner, H. Haffner, F. Schmidt-Kaler, I. L. Chuang, and R. Blatt}}},
\bibinfo{journal}{Nature} \textbf{\bibinfo{volume}{421}},
\bibinfo{pages}{48} (\bibinfo{year}{2003}).
  
\bibitem[{\citenamefont{{N. Timoney, V. Elman, S. Glaser, C. Weiss, M. Johanning, W. Neuhauser, and C. Wunderlich}}(2008)}]{Timoney2008}
\bibinfo{author}{\bibnamefont{{N. Timoney, V. Elman, S. Glaser, C. Weiss, M. Johanning, W. Neuhauser, and C. Wunderlich}}},
\bibinfo{journal}{Phys.\ Rev.\ A.} \textbf{\bibinfo{volume}{77}},
\bibinfo{pages}{052334} (\bibinfo{year}{2008}).

\bibitem[{\citenamefont{{J. J. L. Morton, A. M. Tyryshkin, A. Ardavan, K. Porfyrakis, S. A. Lyon, and G. A. Briggs}}(2005)}]{Morton2005}
\bibinfo{author}{\bibnamefont{{J. J. L. Morton, A. M. Tyryshkin, A. Ardavan, K. Porfyrakis, S. A. Lyon, and G. A. Briggs}}},
\bibinfo{journal}{Phys.\ Rev.\ Lett.} \textbf{\bibinfo{volume}{95}},
\bibinfo{pages}{200501} (\bibinfo{year}{2005}).

\bibitem[{\citenamefont{{E. Collin, G. Ithier, A. Aassime, P. Joyez, D. Vion, and D. Esteve}}(2005)}]{Collin2004}
\bibinfo{author}{\bibnamefont{{E. Collin, G. Ithier, A. Aassime, P. Joyez, D. Vion, and D. Esteve}}},
\bibinfo{journal}{Phys.\ Rev.\ Lett.} \textbf{\bibinfo{volume}{93}},
\bibinfo{pages}{157005} (\bibinfo{year}{2004}).

\bibitem[{\citenamefont{{L. Viola, S. Lloyd, and E. Knill}}(1999)}]{Viola1999}
\bibinfo{author}{\bibnamefont{{L. Viola, S. Lloyd, and E. Knill}}},
\bibinfo{journal}{Phys.\ Rev.\ Lett.} \textbf{\bibinfo{volume}{83}},
\bibinfo{pages}{4888} (\bibinfo{year}{1999}).

\bibitem[{\citenamefont{{L. M. K. Vandersypen, M. Steffen, G. Breyta, C. S. Yannoni, M. H. Sherwood, and I. L. Chuang}}(2001)}]{Vandersypen2001}
\bibinfo{author}{\bibnamefont{{L. M. K. Vandersypen, M. Steffen, G. Breyta, C. S. Yannoni, M. H. Sherwood, and I. L. Chuang}}},
\bibinfo{journal}{Nature} \textbf{\bibinfo{volume}{414}},
\bibinfo{pages}{883} (\bibinfo{year}{2001}).

\bibitem[{\citenamefont{{N. Khaneja, T. Reiss, C. Kehlet, T. Schulte-Herbr\"uggen, and S. J. Glaser}}(2004)}]{Khaneja2004}
\bibinfo{author}{\bibnamefont{{N. Khaneja, T. Reiss, C. Kehlet, T. Schulte-Herbr\"uggen, and S. J. Glaser}}},
\bibinfo{journal}{Jour.\ Mag.\ Res.} \textbf{\bibinfo{volume}{172}},
\bibinfo{pages}{296} (\bibinfo{year}{2004}).

\bibitem[{\citenamefont{{S. Chaudhury, G. A. Smith, K. Schulz, and P. S. Jessen}}(2006)}]{Chaudhury2006}
\bibinfo{author}{\bibnamefont{{S. Chaudhury, G. A. Smith, K. Schulz, and P. S. Jessen}}},
\bibinfo{journal}{Phys.\ Rev.\ Lett.} \textbf{\bibinfo{volume}{96}},
\bibinfo{pages}{043001} (\bibinfo{year}{2006}).

\bibitem[{\citenamefont{{D. Budker, W. Gawlik, D. F. Kimball, S. M. Rochester, v. v. Yashchuck, and A. Weiss}}(2002)}]{Budker2002}
\bibinfo{author}{\bibnamefont{{D. Budker, W. Gawlik, D. F. Kimball, S. M. Rochester, v. v. Yashchuck, and A. Weiss}}},
\bibinfo{journal}{Rev.\ Mod.\ Phys.} \textbf{\bibinfo{volume}{75}},
\bibinfo{pages}{1153} (\bibinfo{year}{2002}).

\bibitem[{\citenamefont{{A. Kuzmich, L. Mandel, N. P. Bigelow}}(2000)}]{Kuzmich2000}
\bibinfo{author}{\bibnamefont{{A. Kuzmich, L. Mandel, N. P. Bigelow}}},
\bibinfo{journal}{Phys.\ Rev.\ Lett.} \textbf{\bibinfo{volume}{85}},
\bibinfo{pages}{1594} (\bibinfo{year}{2000}).

\bibitem[{\citenamefont{{B. Julsgaard, A. Kozhekin, E. S. Polzik}}(2001)}]{Julsgaard2001}
\bibinfo{author}{\bibnamefont{{B. Julsgaard, A. Kozhekin, E. S. Polzik}}},
\bibinfo{journal}{Nature} \textbf{\bibinfo{volume}{413}},
\bibinfo{pages}{6854} (\bibinfo{year}{2001}).

\bibitem[{\citenamefont{{G. A. Smith, S. Chaudhury, A. Silberfarb, I. H. Deutsch, and P. S. Jessen}}(2004)}]{Smith2004}
\bibinfo{author}{\bibnamefont{{G. A. Smith, S. Chaudhury, A. Silberfarb, I. H. Deutsch, and P. S. Jessen}}},
\bibinfo{journal}{Phys.\ Rev.\ Lett.} \textbf{\bibinfo{volume}{93}},
\bibinfo{pages}{163602} (\bibinfo{year}{2004}).

\bibitem[{\citenamefont{{G. Klose, G. A. Smith and P. S. Jessen}}(2001)}]{Klose2001}
\bibinfo{author}{\bibnamefont{{G. Klose, G. A. Smith and P. S. Jessen}}},
\bibinfo{journal}{Phys.\ Rev.\ Lett.} \textbf{\bibinfo{volume}{86}},
\bibinfo{pages}{4721} (\bibinfo{year}{2001}).

\bibitem[{\citenamefont{{H. C. Torrey}}(1949)}]{Torrey1949}
\bibinfo{author}{\bibnamefont{{H. C. Torrey}}},
\bibinfo{journal}{Phys.\ Rev.} \textbf{\bibinfo{volume}{76}},
\bibinfo{pages}{1059} (\bibinfo{year}{1949}).

\bibitem[{\citenamefont{L. Allen and J. H. Eberly}(1987)}]{Allen1987}
\bibinfo{author}{\bibnamefont{{L. Allen and J. H. Eberly}}},
 \textit{\bibinfo{book}{Optical Resonance and Two-Level Atoms}},(\bibinfo{publisher}{Dover Publications,
New York, NY}, \bibinfo{year}{1987}).
 
\bibitem[{\citenamefont{{K. D. Nelson, X. Li and D. S. Weiss}}(2007)}]{Nelson2007}
\bibinfo{author}{\bibnamefont{{K. D. Nelson, X. Li and D. S. Weiss}}},
\bibinfo{journal}{Nature Phys.} \textbf{\bibinfo{volume}{3}},
\bibinfo{pages}{556} (\bibinfo{year}{2007}). 

\bibitem[{\citenamefont{{H. C. N\"agerl, D. Leibfried, H. Rohde, G. Thalhammer, J. Eschner, F. Schmidt-Kaler, and R. Blatt}}(1999)}]{Nagerl1999}
\bibinfo{author}{\bibnamefont{{H. C. N\"agerl, D. Leibfried, H. Rohde, G. Thalhammer, J. Eschner, F. Schmidt-Kaler, and R. Blatt}}},
\bibinfo{journal}{Phys.\ Rev.\ A.} \textbf{\bibinfo{volume}{60}},
\bibinfo{pages}{145} (\bibinfo{year}{1999}).

\bibitem[{\citenamefont{{D. Schrader, I. Dotsenko, M. Khudaverdyan, Y. Miroshnychenko, A. Rauschenbeutel, and D. Meschede}}(2004)}]{Schrader2004}
\bibinfo{author}{\bibnamefont{{D. Schrader, I. Dotsenko, M. Khudaverdyan, Y. Miroshnychenko, A. Rauschenbeutel, and D. Meschede}}},
\bibinfo{journal}{Phys.\ Rev.\ Lett.} \textbf{\bibinfo{volume}{93}},
\bibinfo{pages}{150501} (\bibinfo{year}{2004}).

\bibitem[{\citenamefont{{C. Zhang, S. L. Rolston, and S. Das Sarma}}(2006)}]{Zhang2006}
\bibinfo{author}{\bibnamefont{{C. Zhang, S. L. Rolston, and S. Das Sarma}}},
\bibinfo{journal}{Phys.\ Rev.\ A.} \textbf{\bibinfo{volume}{74}},
\bibinfo{pages}{042316} (\bibinfo{year}{2006}).

\bibitem[{\citenamefont{{I. H. Deutsch}}(2008)}]{Deutsch2008}
\bibinfo{author}{\bibnamefont{{I. H. Deutsch}}}
\bibinfo{journal}{(private communication)}.



\end{thebibliography}
\end{document}